# SUBMODULAR RANK AGGREGATION ON SCORE-BASED PERMUTATIONS FOR DISTRIBUTED AUTOMATIC SPEECH RECOGNITION

*Jun Qi*[1], *Chao-Han Huck Yang*[1], *Javier Tejedor*[2]

[1]Electrical and Computer Engineering, Georgia Institute of Technology, Atlanta, GA, USA
[2] Escuela Politecnica Superior, Universidad San Pablo-CEU, CEU Universities, Madrid, Spain

## ABSTRACT

Distributed automatic speech recognition (ASR) requires to aggregate outputs of distributed deep neural network (DNN)-based models. This work studies the use of submodular functions to design a rank aggregation on score-based permutations, which can be used for distributed ASR systems in both supervised and unsupervised modes. Specifically, we compose an aggregation rank function based on the Lovasz Bregman divergence for setting up linear structured convex and nested structured concave functions. The algorithm is based on stochastic gradient descent (SGD) and can obtain well-trained aggregation models. Our experiments on the distributed ASR system show that the submodular rank aggregation can obtain higher speech recognition accuracy than traditional aggregation methods like Adaboost. Code is available online [1].

## 1. INTRODUCTION

The recent development of computer architectures allows automatic speech recognition (ASR) to be set up in a distributed way [1, 2], where the outputs of different deep neural network (DNN)-based acoustic models are somehow combined to obtain higher ASR accuracy. Our previous work on distributed ASR [3] focused on how to set up a distributed ASR system based on several non-overlapping data subsets, which are produced by applying the submodular data partitioning approach. However, the traditional Adaboost [4] was used in that work to combine the outputs of different subsystems from distributed acoustic models for a better ASR performance. This work proposes a novel method on submodular rank aggregation on score-based permutations for combining scores of acoustic models into a final consistent list of scores.

Rank aggregation is the task of combining a couple of different permutations on the same set of candidates into one ranking list with a better permutation on candidates. More specifically, there are $K$ permutations $\{\mathbf{x}_1, \mathbf{x}_2, ..., \mathbf{x}_K\}$ in total, where the elements $\{X_{1j}, X_{2j}, ..., X_{Nj}\}$, which are in the ranking list $\mathbf{x}_j$, denote either relative orders or numeric values for the candidates. In the framework of rank aggregation on score-based permutations, the elements in the ranking lists represent numerical values and a combined list $\hat{\mathbf{x}}$ with the aggregated scores is used to obtain the relative orders for candidates by sorting the values in $\hat{\mathbf{x}}$.

The Lovasz Bregman (LB) divergence for rank aggregation on the score-based permutations is initially introduced in [5]. The LB divergence is derived from the generalized Bregman divergence parametrized by the Lovasz extension of a submodular function, where a submodular function can be defined via the property of diminishing return [6]. The diminishing return suggests that a function $f : 2^V \to \mathbb{R}_+$ is submodular if $\forall a \in V \setminus B$ and subsets $A \subseteq B \subseteq V$, $f$ satisfies the inequality $f(\{a\} \cup A) - f(A) \geq f(\{a\} \cup B) - f(B)$. Although general discrete optimization problems are NP-hard, the submodular function allows the discrete optimization problems to be efficiently approximated by heuristic algorithms, e.g., an exact minimization and an approximated maximization [7], in a polynomial time with a performance guarantee. Moreover, the Lovasz extension of a submodular function ensures a convex function [8].

[1]https://github.com/uwjunqi/Subrank

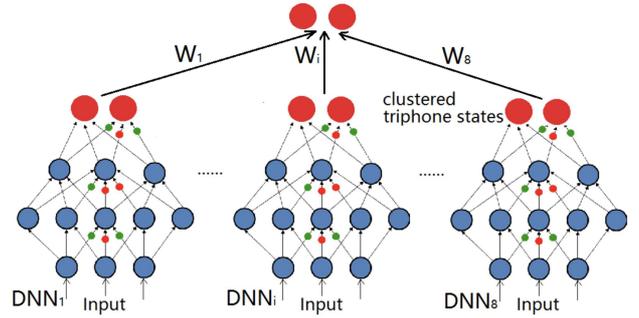

**Fig. 1**. *An illustration of a distributed ASR system.*

The LB divergence to rank aggregation on score-based permutations is briefly discussed in [5, 9], where the LB divergence is capable of measuring the divergence between a score-based permutation and an order-based permutation. This work applies the LB divergence to construct the Normalized Discounted Cumulative Gain (NDCG) [10] loss function for the rank aggregation with an application to the distributed ASR system. The NDCG is a measure of ranking quality to evaluate the effectiveness of ranking algorithms, and it can be taken as an objective function for the rank aggregation of the distributed ASR system.

Figure 1 shows a distributed ASR system, where there are 8 well-trained DNN-based acoustic models and the outputs of all models need to be aggregated to enhance the ASR performance. Our previous work [3, 11] focused on how to use submodular data partitioning and selecting algorithms to split the full dataset into 8 different non-overlapping subsets, and made use of the subsets to train 8 different DNN-based acoustic models. Then, the scoring lists of 8 acoustic models are aggregated to generate a consensus scoring list based on submodular rank aggregation in this work.

Our contribution in this work includes: (1) the NDCG loss function is constructed based on the LB divergence; (2) linear and Nested structured frameworks are formulated for constructing optimization problems of the submodular rank aggregation; (3) efficient learning and inference algorithms are proposed; (4) the algorithms are applied to a distributed ASR system.

The remainder of the paper is organized as follows: Section 2 introduces the submodular function based on the LB divergence and the related NDCG loss function, and then the linear and nested rank aggregation functions are discussed in Section 3. The experimental results of the submodular rank aggregation for the distributed ASR system are shown in Section 4 and the paper is concluded in Section 5.

## 2. LB DIVERGENCE AND THE NDCG LOSS FUNCTION

The LB divergence as a utility function was firstly proposed in [12], and the related definitions for the LB divergence are briefly summarized in **Definitions** 1, 2, 3.

**Definition 1.** For a submodular function $f$ and an order-based permutation $\sigma$, then there is a chain of sets $\phi = S_0^\sigma \subseteq S_1^\sigma \subseteq ... \subseteq S_N^\sigma = V$, with which the vector $h_\sigma^f$ is defined as (1).

$$h_\sigma^f(\sigma(i)) = f(S_i^\sigma) - f(S_{i-1}^\sigma), \forall i = 1, 2, ..., |V| \quad (1)$$

where $V$ is the ground set with all elements, and $|V|$ is the number of candidates.

**Definition 2.** For a submodular function $f$ and a score-based permutation $\mathbf{x}$, we define a permutation $\sigma_\mathbf{x}$ such that $\mathbf{x}[\sigma_\mathbf{x}(1)] \geq \mathbf{x}[\sigma_\mathbf{x}(2)] \geq ... \geq \mathbf{x}[\sigma_\mathbf{x}(N)]$, and a chain of sets $\phi = S_0^{\sigma_\mathbf{x}} \subseteq S_1^{\sigma_\mathbf{x}} \subseteq ... \subseteq S_N^{\sigma_\mathbf{x}} = V$, with which the vector $h_{\sigma_\mathbf{x}}^f$ is defined as (2).

$$h_{\sigma_\mathbf{x}}^f(\sigma_\mathbf{x}(i)) = f(S_i^{\sigma_\mathbf{x}}) - f(S_{i-1}^{\sigma_\mathbf{x}}), \forall i = 1, 2, ..., N \quad (2)$$

**Definition 3.** Given a submodular function $f$ and its associated Lovasz extension $\hat{f}$, we define the LB divergence $d_{\hat{f}}(\mathbf{x}||\sigma)$ for measuring the divergence of permutations between a score-based permutation $\mathbf{x}$ and an order-based permutation $\sigma$. The LB divergence is defined as (3).

$$d_{\hat{f}}(\mathbf{x}||\sigma) = <\mathbf{x}, h_{\sigma_\mathbf{x}}^f - h_\sigma^f> \quad (3)$$

where both $h_\sigma^f$ and $h_{\sigma_\mathbf{x}}^f$ are defined via definitions 1 and 2, respectively.

Next, we study how to apply the LB divergence to obtain the NDCG loss function, which is used in the rank aggregation on score-based permutations.

**Definition 4.** Given an order-based permutation $\sigma$ with candidates from a ground set $V = \{1, 2, ..., N\}$, and a set of relevance scores $\{r(1), r(2), ..., r(N)\}$ associated with $\sigma$, the NDCG score for $\sigma$ is defined as (4), where $D(\cdot)$ denotes a discounted term and $\pi$ denotes the ground truth. Accordingly, the NDCG loss function $L(\sigma)$ is defined as (5).

$$NDCG(\sigma) = \frac{1}{\sum_{i=1}^{N} r(\pi(i))D(i)} \sum_{i=1}^{N} r(\sigma(i))D(i) \quad (4)$$

$$L(\sigma) = 1 - NDCG(\sigma) = \frac{\sum_{i=1}^{N} r(\pi(i))D(i) - r(\sigma(i))D(i)}{\sum_{i=1}^{N} r(\pi(i))D(i)} \quad (5)$$

**Corollary 1.** For a submodular function $f(X) = g(|X|)$, the LB divergence $d_{\hat{f}}(\mathbf{x}||\sigma)$ associated with $f$ is derived as (6).

$$d_{\hat{f}}(\mathbf{x}||\sigma) = \sum_{i=1}^{N} \mathbf{x}(\sigma_\mathbf{x}(i))\delta_g(i) - \sum_{i=1}^{N} \mathbf{x}(\sigma(i))\delta_g(i) \quad (6)$$

where $\delta_g(i) = g(i) - g(i-1)$, $g(\cdot)$ is a concave function, $|X|$ denotes the cardinality of the set $X$, and $\sigma_\mathbf{x}$ and $\sigma$ refer to the score-based and order-based permutations, respectively.

It is found that $d_{\hat{f}}(\mathbf{x}||\sigma) \propto L(\sigma)$, and the normalized $d_{\hat{f}}(\mathbf{x}||\sigma)$ can be applied as a utility function for the NDCG loss measurement because the normalization term in (5) is constant. Besides, the LB divergence guarantees an upper bound for the NDCG loss function as shown in Proposition 1.

**Proposition 1.** Given a score-based permutation $\mathbf{x}$ and a concave function $g$, the LB divergence $d_{\hat{f}}(\mathbf{x}||\sigma)$ defined as (6) provides a constant upper bound to the NDCG loss function. Specifically,

$$L(\sigma) \leq \frac{n}{Z} \cdot \epsilon \cdot (g(1) - g(|V|) + g(|V|-1))$$
$$\leq \frac{\epsilon \cdot (g(1) - g(|V|) + g(|V|-1))}{\min_i \mathbf{x}(\sigma_\mathbf{x}(i))\delta_g(i)} \quad (7)$$

where $n$ is the number of permutation $\mathbf{x}$, $\epsilon = \max_{i,j} |\mathbf{x}(i) - \mathbf{x}(j)|$, $Z = \sum_{i=1}^{n} \mathbf{x}(\sigma_\mathbf{x}(i))\delta_g(i)$ is a normalization term, and the upper bound for $L(\sigma)$ is independent of the permutation variable $\sigma$.

*Proof.* To obtain (7), we firstly show that, given a monotone submodular function $f$ and any permutation $\sigma$, there is an inequality (8) for $d_{\hat{f}}(\mathbf{x}||\sigma)$:

$$d_{\hat{f}}(\mathbf{x}||\sigma) \leq \epsilon n \cdot (\max_j f(j) - \min_j f(j|V\backslash\{j\})) \quad (8)$$

where $\epsilon = \max_{i,j} |\mathbf{x}(i) - \mathbf{x}(j)|$, and $f(j|V\backslash\{j\}) = f(V) - f(V\backslash\{j\})$ is a marginal gain. Furthermore, by setting $f(X) = g(|X|)$, we can obtain $\max_j f(j) = g(1)$, and by applying the submodular diminishing return property $\min_j f(j|V\backslash\{j\}) = f(V) - f(V\backslash\{j\}) = g(|V|) - g(|V|-1)$. Thus, by setting $L(\sigma) = \frac{1}{Z} d_{\hat{f}}(\mathbf{x}||\sigma)$, and noting that $Z \geq n \cdot \min_i \sigma_\mathbf{x}(i)\delta_g(i)$, the proof for (7) is completed. □

## 3. THE LEARNING FRAMEWORKS FOR SUBMODULAR RANK AGGREGATION

Section 3 discusses two learning frameworks based on the NDCG loss function as introduced in Section 2, and the associated algorithms are proposed accordingly.

### 3.1. The Linear Structured Framework

The first learning framework based on the LB divergence is built upon a linear structure. Suppose that there are $|Q|$ queries and $K$ score-based permutations $\{\mathbf{x}_1^q, \mathbf{x}_2^q, ..., \mathbf{x}_K^q\}$ associated with the query $q \in Q$, $\forall q \in Q$, and $\pi^q \in \{1, 2, ..., N\}$. The LB divergence $d_{\hat{f}}(\mathbf{x}_i^q||\pi^q)$ computes the divergence between $\mathbf{x}_i^q$ and $\pi^q$. The problem focuses on how to assign the weight vector $\mathbf{w} = (w_1, w_2, ..., w_K)^T$ to $d_{\hat{f}}(\mathbf{x}||\pi^q) = (d_{\hat{f}}(\mathbf{x}_1^q||\pi^q), ..., d_{\hat{f}}(\mathbf{x}_K^q||\pi^q))^T$. The objective function of the linear structure is formulated in (9), where $\lambda$ denotes a regularization term.

**Algorithm 1** Inference Algorithm 1
1. Input: a test data $\{q, N, K, X\}$, and the trained weights $\mathbf{w}^* = \{w_1^*, w_2^*, ..., w_K^*\}$.
2. Compute $R_X = \sum_{i=1}^{K} w_i^* \mathbf{x}_i$.
3. Argsort $R_X$ in a decreasing order $\rightarrow \hat{\sigma}$.
4. Output: $\hat{\sigma}$ and the sorted scores $R_X(\hat{\sigma})$.

$$\min_{\mathbf{w}} \frac{1}{|Q|} \sum_{q \in Q} \sum_{i=1}^{K} w_i d_{\hat{f}}(\mathbf{x}_i^q || \pi^q) + \frac{\lambda}{2} \sum_{i=1}^{K} w_i^2 \quad (9)$$
$$s.t., \sum_{i=1}^{K} w_i = 1, \ w_i \geq 0$$

The stochastic gradient descent (SGD) algorithm is applied to obtain $\nabla^{sgd}$ as (10), and update the parameter $\mathbf{w}$ based on (11) to satisfy the constraints in (9).

$$\nabla_i = \frac{1}{|Q|} \sum_{q \in Q} d_{\hat{f}}(x_i^q || \pi^q) + \lambda w_i \quad (10)$$

$$w_i^{(t+1)} = \frac{w_i^{(t)} \exp(-\mu \nabla_i^{sgd})}{\sum_{j=1}^{K} w_j^{(t)} \exp(-\mu \nabla_j^{sgd})} \quad (11)$$

The update of (11) goes through all $i = 1, ..., K$ and each query $q \in Q$. Several iterations are repeated until $\mathbf{w}$ gets convergence.

**Proposition 2.** In the linear structured framework, given the number of candidates $N$ and the number of permutations $K$, the computational complexity in the training stage is $O(NK)$.

The inference process is formulated as follows: given a test data $\{q, N, K, X\}$, where $q$ is a query, $N$ and $K$ have been defined as above, and $X = \{\mathbf{x}_1^q, ..., \mathbf{x}_K^q\}$ represents the score-based permutation associated with the query $q$, we estimate an optimal order-based permutation $\hat{\sigma}$ as defined in (12).

$$\hat{\sigma} = \arg\min_{\pi^q} \sum_{i=1}^{K} w_i^* d_{\hat{f}}(\mathbf{x}_i^q || \pi^q) \quad (12)$$

where $w^*$ refers to the weight vector trained in the training stage. Generally, the problem in (12) is an NP-hard combinatorial problem, but the LB divergence provides a close-form solution $\hat{\sigma} = \sigma_\mu$, where $\mu = \sum_{i=1}^{K} w_i^* \mathbf{x}_i^q$. The complete inference algorithm is shown in Algorithm 1. Note that the order-based permutation $\hat{\sigma}$ is finally obtained by sorting the numerical values in $R_X$ and the permutation associated with $R_X$ is returned in a decreasing order.

### 3.2. The Nested Structured Framework

The linear structured framework involves several potential problems: the first problem is that the score-based permutations $\{\mathbf{x}_i\}_{i=1}^{K}$ may not interact with each other, since one permutation might be partially redundant with another; the second problem lies in the fact that a permutation $\mathbf{x}_i$ tends to become dominant over the rest. To overcome these problems, an additional hidden layer is utilized to construct a nested structured framework.

The objective function for the nested structured framework is formulated as (13), where $K_1$ and $K_2$ separately represent the numbers of units in the input and hidden layers, $\mathbf{W}_1 \in R^{K_2 \times K_1}$ and $\mathbf{W}_2 \in R^{1 \times K_2}$ denote weights for the bottom and upper layers, respectively, and $\lambda_1$ and $\lambda_2$ refer to regularization terms. By setting both $\Phi_1$ and $\Phi_2$ as increasing concave functions, the objective function becomes a concave function and thus it needs to be maximized with respect to $\mathbf{W}_1$ and $\mathbf{W}_2$.

$$\max_{W_1, W_2} \frac{1}{|Q|} \sum_{q \in Q} \Phi_2(\sum_{i=1}^{K_2} W_2(i) \Phi_1(\sum_{j=1}^{K_1} W_1(i,j) d_{\hat{f}}(\mathbf{x}_j^q || \pi^q)))$$
$$+ \frac{\lambda_1}{2} ||W_1||_F^2 + \frac{\lambda_2}{2} ||W_2||_F^2 \quad (13)$$
$$s.t., \sum_{j=1}^{K_1} W_1(i,j) = 1, \sum_{i=1}^{K_2} W_2(i) = 1,$$
$$W_1(i,j) \geq 0, \ W_2(i) \geq 0, \ \forall i,j$$

The update for weights $\mathbf{W}_1$ and $\mathbf{W}_2$ follows a feed-forward manner that is similar to that employed for a standard Multiple Layer Perceptron (MLP) training.

As to the update for the weights of the bottom layer, the temporal variables $\delta_1(i)$ and $\nabla_1(i,j)$ need to be firstly computed by (14) and (15), and $\mathbf{W}_1$ is updated by (16), where $\mu$ refers to the learning rate.

$$\delta_1(i) = \sum_{j=1}^{K_1} W_1^{(t)}(i,j) d_{\hat{f}}(\mathbf{x}_j^q || \pi^q) \quad (14)$$

$$\nabla_1(i,j) = (\Phi_1(\delta_1(i)))' \sum_{j=1}^{K_1} d_{\hat{f}}(\mathbf{x}_j^q || \pi^q) + \lambda_1 W_1^{(t)}(i,j) \quad (15)$$

$$W_1^{(t+1)}(i,j) = \frac{W_1^{(t)}(i,j) \exp(-\mu \nabla_1(i,j))}{\sum_{j=1}^{K_1} W_1^{(t)}(i,j) \exp(-\mu \nabla_1(i,j))} \quad (16)$$

The procedure of updating $\mathbf{W}_2$ starts as soon as the update of $\mathbf{W}_1$ is done. The new $\delta_2$ and $\nabla_2(i)$ are separately derived by (17) and (18) based on the message propagation from the bottom layer. Finally, the update for $\mathbf{W}_2$ is conducted by (19).

$$\delta_2 = \sum_{i=1}^{K_2} W_2^{(t)}(i) \Phi_1(\delta_i(i)) \quad (17)$$

$$\nabla_2(i) = (\Phi_2(\delta_2))' \Phi_1(\delta_1(i)) + \lambda_2 W_2^{(t)}(i) \quad (18)$$

$$W_2^{(t+1)}(i) = \frac{W_2^{(t)}(i) \exp(-\mu \nabla_2(i))}{\sum_{j=1}^{K_2} W_2^{(t)}(j) \exp(-\mu \nabla_2(j))}, \ \forall i \quad (19)$$

**Proposition 3.** In the nested structured framework, given the numbers $K_1$ and $K_2$ for the input and hidden layers, respectively, and the number of candidates $N$, the computational complexity for the entire training process is $O(NK_1 + K_1 K_2)$.

The inference for the nested structured framework shares the same steps that the linear structured framework except that the step 2 in Algorithm 1 is replaced by (20).

$$R_X = \Phi_2(\sum_{i=1}^{K_2} W_2(i) \Phi_1(\sum_{j=1}^{K_1} W_1(i,j) \mathbf{x}_j^q)) \quad (20)$$

## 4. EXPERIMENTS

This section presents the experimental setups and empirical results on the distributed ASR system. Our distributed ASR system is shown in Figure 1, where there are 8 DNN-based acoustic models and each DNN shares the same initial setup in terms of the model architectures and ReLU activation functions. A full training dataset is partitioned into 8 non-overlapping subsets by means of the robust submodular data partitioning algorithm [3], and each subset is employed to train a particular DNN-based acoustic model, which results in 8 different DNN-based acoustic models. The 8 lists of acoustic scores from the acoustic models need to be aggregated into a final list with a higher ASR accuracy. Instead of adopting the aggregation method based on Adaboost of our previous work [3], we apply the approach of submodular rank aggregation discussed in this work.

### 4.1. Experimental setup

Following the steps in [3], the submodular partitioning functions were composed based on the prior phonetic knowledge that a triphone corresponds to 8 different biphones based on the phonetic knowledge including 'Place of Articulation', 'Production Manner', 'Voicedness' and 'Miscellaneous'. For training each DNN-based acoustic model, the full dataset is split into 8 disjoint data subsets by formulating the problem as a robust submodular data partitioning algorithm in [3].

Our experiments were conducted on the 1300 hours of conversational English telephone speech data from the Switchboard, Switchboard Cellular, and Fisher databases. The development and testing datasets were the 2001 and 2002 NIST Rich Transcription development sets, with 2.2 hours and 6.3 hours of acoustic data, respectively. Data preprocessing includes extracting 39-dimensional Mel Frequency Cepstrum Coefficient (MFCC) features that correspond to 25.6ms of speech signals. Besides, mean and variance speaker normalization were also applied.

The acoustic models are initialized as clustered tri-phones modeled by 3-state left-to-right hidden Markov models (HMMs) [13]. The state emission probability in the HMMs was modeled from the Gaussian mixture model (GMM). The DNN targets consisted of approximately 7800 clustered tri-phone states [14]. All sequential labels corresponding to the training data were generated by forced-alignment based on HMM-GMM. A 3-gram language model, built from the training materials, was used for the decoding.

The units at the input layer of each of the 8 DNNs correspond to a long-context feature vector that was generated by concatenating 11 consecutive frames of the primary MFCC feature followed by a discrete cosine transformation (DCT) [15, 16]. Thus, the dimension of the initial long-context feature was 429 and the number was reduced to 361 after DCT. There were 4 hidden layers in total with a setup of 1024-1024-1024-1024 for the DNN construction [17]. The parameters of the hidden layers were initialized via Restricted Boltzman Machine pre-training [18], and the fine-tuned by the Multi-layer Perceptron Back-propagation algorithm. Besides, the feature-based maximum likelihood linear regression was applied to the DNN speaker adaption [19].

When the training process of all the DNN-based acoustic models was done, the final posteriors of the clustered tri-phones associated with the training data should be separately obtained from each of the DNN-based acoustic models. Those posteriors were taken as permutation data for training the submodular rank aggregation models. In the testing stage, the posteriors collected from the 8 DNN-based acoustic models are combined to one permutation that is expected to be as close to the ground truth as possible. As for the configuration of the submodular rank aggregation formulations, the number of permutations $K$ is set to 8, and the dimension of a permutation is configured as 7800, which matches the number of clustered tri-phone states. Besides, the number of units of the hidden layer in the nested structured framework is set to 120, and there is only one output corresponding to the final aggregation permutation.

### 4.2. Experimental results

Table 1 shows the ASR decoding results from each of the DNN-based acoustic models, and Table 2 represents the combined ones based on the different aggregation methods. The results suggest that the submodular rank aggregation methods based on the linear and nested structured formulation obtain a better result than the baseline system based on the Adaboost approach and a simple averaging method. The marginal gain obtained with the nested structured formulation with regards to the linear structured one arises from the potential bias caused by the forced-alignment of the clustered tri-phones.

| Model | DNN1 | DNN2 | DNN3 | DNN4 |
|-------|------|------|------|------|
| PER   | 28.3 | 28.7 | 29.3 | 28.6 |
| Model | DNN5 | DNN6 | DNN7 | DNN8 |
| PER   | 28.8 | 28.7 | 28.6 | 28.4 |

**Table 1**. Phone Error Rate (PER) (%).

| Method | PER |
|--------|-----|
| Averaging | 28.1 |
| Adaboost | 27.4 |
| Linear-LBD | 26.8 |
| Nested-LBD | **26.4** |

**Table 2**. Phone Error Rate (PER) (%). 'LBD' stands for the Lovasz Bregman divergence.

## 5. CONCLUSIONS

This work proposes a submodular rank aggregation on score-based permutations based on the LB divergence in both linear and nested structured frameworks for a distributed ASR system. Besides, the related learning algorithms based on SGD are used to ensure optimal solutions to the submodular rank aggregation problem. Our experiments on the distributed ASR system show that the nested LB divergence can obtain significantly more gains. However, the gains obtained with respect to the other approaches are lower to varying degrees. In addition, our methods can be scalable to large-scale datasets because of their low computational complexity in the inference procedure.

Future work will study how to generalize the nested structure to a deeper one with more hidden layers. Although the convexity/concavity of the objective function with a deep structure can be maintained, the use of the message-passing to deeper layers cannot obtain a satisfying result on the task of distributed ASR. Thus, an improved learning algorithm for training an LB-divergence based objective function with a deeper structure is necessary.

# 6. REFERENCES

[1] Guangsen Wang and Khe Chai Sim, "Regression-based context-dependent modeling of deep neural networks for speech recognition," *IEEE/ACM Transactions on Audio, Speech and Language Processing (TASLP)*, vol. 22, no. 11, pp. 1660–1669, 2014.

[2] Kadri Hacioglu and Bryan Pellom, "A distributed architecture for robust automatic speech recognition," in *2003 IEEE International Conference on Acoustics, Speech, and Signal Processing (ICASSP)*. IEEE, 2003, vol. 1, pp. I–I.

[3] Jun Qi and Javier Tejedor, "Robust submodular data partitioning for distributed speech recognition," in *2016 IEEE International Conference on Acoustics, Speech and Signal Processing (ICASSP)*. IEEE, 2016, pp. 2254–2258.

[4] Gunnar Rätsch, Takashi Onoda, and K-R Müller, "Soft margins for adaboost," *Machine Learning*, vol. 42, no. 3, pp. 287–320, 2001.

[5] Rishabh Iyer and Jeff Bilmes, "The lovász-bregman divergence and connections to rank aggregation, clustering, and web ranking," *arXiv preprint arXiv:1308.5275*, 2013.

[6] Jack Edmonds, "Submodular functions, matroids, and certain polyhedra," in *Combinatorial Optimization—Eureka*, pp. 11–26. Springer, 2003.

[7] Satoru Fujishige, *Submodular functions and optimization*, vol. 58, Elsevier, 2005.

[8] László Lovász, "Submodular functions and convexity," in *Mathematical Programming The State of the Art*, pp. 235–257. Springer, 1983.

[9] Jun Qi, Xu Liu, Javier Tejedor, and Shunsuke Kamijo, "Unsupervised submodular rank aggregation on score-based permutations," *arXiv preprint arXiv:1707.01166*, 2017.

[10] Kalervo Järvelin and Jaana Kekäläinen, "Cumulated gain-based evaluation of ir techniques," *ACM Transactions on Information Systems (TOIS)*, vol. 20, no. 4, pp. 422–446, 2002.

[11] Jun Qi, Xu Liu, Shunshuke Kamijo, and Javier Tejedor, "Distributed submodular maximization for large vocabulary continuous speech recognition," in *2018 IEEE International Conference on Acoustics, Speech and Signal Processing (ICASSP)*. IEEE, 2018, pp. 2501–2505.

[12] Rishabh Iyer and Jeff Bilmes, "Submodular bregman divergences with applications," *Neural Information Processing Society (NIPS)*, pp. 2933–2941, 2012.

[13] Sean R Eddy, "Hidden markov models," *Current opinion in structural biology*, vol. 6, no. 3, pp. 361–365, 1996.

[14] Steve J Young, Julian J Odell, and Philip C Woodland, "Tree-based state tying for high accuracy acoustic modelling," in *Proceedings of the workshop on Human Language Technology*. Association for Computational Linguistics, 1994, pp. 307–312.

[15] J. Qi, D. Wang, and J. Tejedor Noguerales, "Subspace models for bottleneck features," in *Interspeech*. ISCA, 2013, pp. 1746–1750.

[16] J. Qi, D. Wang, and and J. Tejedor Noguerales J. Xu, "Bottleneck features based on gammatone frequency cepstral coefficients," in *Interspeech*. ISCA, 2013, pp. 1751–1755.

[17] J. Qi, J. Du, S.M. Siniscalchi, and C.-H. Lee, "A theory on deep neural network based vector-to-vector regression with an illustration of its expressive power in speech enhancement," *IEEE/ACM Transactions on Audio, Speech, and Language Processing (TASLP)*, vol. 27, no. 12, pp. 1932–1943, 2019.

[18] Geoffrey E Hinton and Ruslan R Salakhutdinov, "Reducing the dimensionality of data with neural networks," *science*, vol. 313, no. 5786, pp. 504–507, 2006.

[19] Mark JF Gales, "Maximum likelihood linear transformations for hmm-based speech recognition," *Computer speech & language*, vol. 12, no. 2, pp. 75–98, 1998.